\documentclass[aps,prb,amsmath,reprint,superscriptaddress]{revtex4-1}
\usepackage{amsmath,txfonts,bm,color,dcolumn,graphicx,multirow}
\begin{document}

\title{Zero-Field Splitting Parameters from Four-Component Relativistic Methods}

\author{Ryan D. Reynolds}
\email{ryanreynolds2018@u.northwestern.edu}
\affiliation{Department of Chemistry, Northwestern University, 2145 Sheridan Rd., Evanston, IL 60208, USA.}
\author{Toru Shiozaki}
\affiliation{Department of Chemistry, Northwestern University, 2145 Sheridan Rd., Evanston, IL 60208, USA.}
\date{\today}

\begin{abstract}
We report an approach for determination of zero-field splitting parameters from four-component relativistic calculations.  
Our approach involves neither perturbative treatment of spin--orbit interaction nor truncation of the spin--orbit coupled states. 
We make use of a multi-state implementation of relativistic complete active space perturbation theory (CASPT2),
partially contracted $N$-electron valence perturbation theory (NEVPT2), and multi-reference configuration interaction theory (MRCI), all with the 
fully internally contracted \emph{ansatz}.  
A mapping is performed from the Dirac Hamiltonian to the pseudospin Hamiltonian, using correlated energies and the magnetic moment 
matrix elements of the reference wavefunctions.  Direct spin--spin coupling is naturally included through the full 2-electron Breit interaction.
Benchmark calculations on chalcogen diatomics and pseudotetrahedral cobalt(II) complexes show accuracy comparable 
to the commonly used state-interaction with spin--orbit (SI-SO) 
approach, while tests on a uranium(III) single-ion magnet suggest that for actinide complexes 
the strengths of our approach through the more robust treatment of spin--orbit effects 
and the avoidence of state truncation are of greater importance.  
\end{abstract}

\maketitle
\section{Introduction}

The past two decades have witnessed a continuous growth of interest in molecular magnets due to their potential for use in 
quantum information processing, high-density information storage, and spintronics.\cite{leuenberger01nat,bogani08nm,woodruff13cr}  
The magnetic properties of open-shell complexes are conventionally modeled using a phenomenological spin Hamiltonian, whose 
parameters are determined experimentally via electron paramagnetic resonance spectroscopy, magnetometry, inelastic neutron 
scattering, and other methods.\cite{abragam70,baker15sb,krzystek16dt}
To second order, the spin Hamiltonian is expressed as 
\begin{align}
    \hat{H} &= D \left( \hat{S}_z^2 - \frac{1}{3} S(S+1) \right) + E\left( \hat{S}_x^2 - \hat{S}_y^2 \right) + \mu_B \mathbf{B} \cdot \mathbf{g} \cdot \hat{\mathbf{S}}
\end{align}
with axial and rhombic zero-field splitting (ZFS) parameters $D$ and $E$, the Bohr magneton $\mu_B$, external magnetic field $\mathbf{B}$, the 
$g$-matrix $\mathbf{g}$ expressing the strength and anisotropy of the electronic Zeeman effect, and spin operator $\hat{\mathbf{S}}$.  
Additional terms can be added for smaller effects such as hyperfine coupling, interaction with the nuclear quadrupole moment, 
higher-order ZFS contributions, and so on.  
One of the key parameters necessary for single-ion magnetism is easy axis ZFS, which contributes to the energy barrier to 
reversal of spin magnetization.  
The ZFS is a breaking of degeneracy of the ground spin multiplet of an open-shell complex with total spin of 1 or greater, 
which results from spin--orbit and spin--spin coupling, and its 
accurate determination is essential for an understanding of the behavior of molecular magnets.  

Several methods are already available for the prediction of spin Hamiltonian parameters.  
The anisotropic $g$-matrix has been predicted using both relativistic density functional theory and wavefunction-based methods.  
\cite{malkin05jcp,bolvin06cpc,vancoillie07cpc,neese07epr,repisky10cpl,chibotaru12jcp,ganyushin13jcp,gohr15jpca}
For the prediction of zero-field energy splittings, the state-interaction with spin--orbit (SI-SO) approach has been widely used.\cite{ganyushin06jcp}
Van den Heuvel \emph{et al}.\cite{vandenheuvel16pccp} have introduced a simplified approach where
configuration-averaged Hartree--Fock orbitals are used in a complete active space configuration interaction (CASCI) diagonalization with spin--orbit effects included.  
This gives comparable results to complete active space self-consistent field (CASSCF) followed by SI-SO for 
lanthanide systems with weak crystal field effects, although neither method could accurately reproduce the experimental low-energy excitation spectra.\cite{jiang10acie} 
Contributions from direct spin--spin coupling have been introduced in the context of density functional theory (DFT).\cite{neese06jacs}

In this work, we demonstrate a robust method for the \emph{ab initio} determination of ZFS parameters of
mononuclear complexes with strong spin--orbit and spin--spin coupling.  
Our approach makes use of four-component relativistic CASSCF,\cite{jensen96jcp,bates15jcp,reynolds18jcp} which can be extended with 
perturbation theory for dynamical correlation\cite{shiozaki15jctc,zhang18jctc} or the Gaunt or full Breit interaction\cite{kelley13jcp}
to include direct spin--spin coupling.  
To extract spin Hamiltonian parameters from the \emph{ab initio} energies and wavefunctions, we 
have adapted the pseudospin mapping approach of Chibotaru and Ungur\cite{chibotaru12jcp} for the four-component wavefunctions used here.  

We begin by reviewing the relevant theory for density fitted, 4-component relativistic CASSCF, fully internally contracted complete active 
space perturbation theory (CASPT2), and partially contracted $N$-electron valence perturbation theory (NEVPT2).  
We also review the procedure for mapping to the pseudospin Hamiltonian before going on to benchmark the methods for several series of molecules.  
We start with 
a series of chalcogen diatomics, comparing 
to existing methods and examining the importance of multi-state approaches in each case.  We then present example calculations of ZFS 
parameters for single-ion magnets with moderate and very strong spin--orbit effects.  
We begin with the pseudotetrahedral cobalt complexes [Co($X$Ph)$_4^{2-}$ for $X$ = O, S, and Se,\cite{zadrozny11jacs,zadrozny13poly,suturina15ic,suturina17ic},
then move on to the actinide complex U(H$_2$BPz$_2$)$_3$,\cite{rinehart10jacs,meihaus11ic,baldovi13cs,spivak17jpca}
where spin--orbit effects are much stronger and should ideally be included concurrently with orbital optimization.  

\section{Theory}

\subsection{Relativistic CASSCF}

Throughout this section, we use $I$, $J$, and $K$ to label Slater determinants, while $L$, $M$, and $N$ refer to
CASSCF states.  The molecular orbital indices $i$ and $j$ refer to closed orbitals, $r$, $s$, and $t$ to active orbitals, 
$a$ and $b$ to virtual orbitals; $w$, $x$, $y$, and $z$ are used as generic orbital indices, and $\mu$ is used for atomic basis functions.   
Our four-component molecular spinors are constructed from a basis set of 2-spinor atomic basis functions 
generated using restricted kinetic balance (RKB).\cite{stanton84jcp,dyall90cpl}
Starting orbitals are generated from a converged Dirac--Hartree--Fock wavefunction,\cite{kelley13jcp} 
with active electrons removed when appropriate.

At the level of CASSCF, the multi-configuration wavefunction is parametrized as
\begin{align}
    |M\rangle &= \sum_I c_{I,M} |I\rangle
,\end{align}
where we seek to find eigenfunctions of the Dirac Hamiltonian within the CASCI space:
\begin{align}
    \langle M | \hat{H} | N \rangle &= \delta_{MN} E_M
.\end{align}
The relativistic Dirac--Breit Hamiltonian can be expressed in second quantized form as 
\begin{align}
    &\hat{H} = \sum_{xy}h_{xy}\hat{E}_{xy} + \frac{1}{2}\sum_{xyzw}v_{xy,zw}\hat{E}_{xy_,zw}
\\  &\hat{h}(1) = c^2(\beta - I_4) + c(\boldsymbol{\alpha}_1\cdot\hat{\mathbf{p}}_1) - \sum_{A}^\text{atoms}\frac{Z_A}{r_{1A}}\mathrm{erf}\left(\sqrt{\zeta_A}r_{iA}\right)
\\  &\hat{v}(1,2) = \frac{1}{r_{12}} - \frac{\boldsymbol{\alpha}_1\cdot\boldsymbol{\alpha}_2}{r_{12}}
                  - \frac{   (\boldsymbol{\alpha}_1\cdot\boldsymbol{\nabla}_1)  (\boldsymbol{\alpha}_2\cdot\boldsymbol{\nabla}_2)  r_{12} }{2}
                  \label{v12}
,\end{align}
with the excitation operators defined
\begin{align}
    &\hat{E}_{xy} = \hat{a}_x^\dagger \hat{a}_y
\\  &\hat{E}_{xy,zw} = \hat{a}_x^\dagger \hat{a}_z^\dagger \hat{a}_w \hat{a}_y
,\end{align}
and the nuclear potential modeled with a Gaussian charge distribution with exponent $\zeta_A$ and total charge $Z_A$.\cite{visscher97adndt}
We can also obtain the Dirac--Coulomb--Gaunt or Dirac--Coulomb Hamiltonian
by neglecting the last one or two terms (respectively) of Eq.~\eqref{v12}.\cite{kelley13jcp}

We perform a two-step optimization procedure to obtain the CASSCF wave functions, 
where for each macroiteration the state-specific energies are made stationary with respect to $c_{I,M}$ coefficients, \emph{i.e.},
\begin{align}
    \frac{\partial E_M}{\partial c_{I,M}} &= 0
.\end{align}
The $c_{I,M}$ coefficients are obtained using the relativistic analogue of full configuration interaction,
described previously by one of the authors.\cite{jensen96jcp,bates15jcp}  
The molecular coefficients are optimized through a series of macroiterations, 
which makes the state-averaged energy stationary with respect to orbital coefficients $C_{\mu p}$ with appropriate normalizations,
\begin{align}
    &\mathbf{C} = \mathrm{exp}(-\boldsymbol{\kappa})
,\\ &\sum_M \frac{\partial E_M}{\partial \kappa_{pq}} = 0
.\end{align}
The orbital optimization is performed using a second-order minimax algorithm, in which explicit construction of the orbital Hessian matrix is avoided by
direct multiplication of density fittted integrals to trial vectors; further details are given in Ref.~\onlinecite{reynolds18jcp}.

\subsection{Relativistic XMS-CASPT2 and NEVPT2}

For the relativistic dynamical correlation theories, we employ a fully internally contracted (FIC) \emph{ansatz}
for the first-order wavefunction, using either the multi-state, multi-reference (MS-MR) parametrization:
\begin{align}
    &|\Psi_M\rangle = \sum_N \hat{T}_{MN} |N\rangle
\end{align}
or the single-state, single-reference (SS-SR) parametrization:
\begin{align}
    &|\Psi_M\rangle = \hat{T}_{MM} |M\rangle
.\end{align}
The former choice comes at higher cost but ensures the invariance of the correlated wavefunction with respect to rotation among reference states.  

For post-CASSCF inclusion of dynamical correlation, we make use of the extended multi-state (XMS)
algorithm for quasi-degenerate perturbation theory proposed by Granovsky.\cite{granovsky11jcp,vlaisavljevich16jctc}
This is necessary to ensure invariance among rotations between reference states, and can be achieved by 
diagonalizing the zeroth-order Hamiltonian in the basis of reference states:
\begin{align}
    &\sum_{M} \langle L | \hat{H}_0 | M \rangle U_{MN} = U_{LN} \tilde{E}_N
\\  &|\tilde{M}\rangle = \sum_N |N\rangle U_{NM}
.\end{align}
Using the rotated reference functions $|\tilde{M}\rangle$, the implementation of XMS-CASPT2
is the same as conventional MS-CASPT2.  
There is no need for an extended multi-state algorithm for NEVPT2, because in this case the zeroth-order Hamiltonian is already diagonal.   

The zeroth-order Hamiltonian for (X)MS-CASPT2 is given by the state-averaged generalized Fock operator:
\begin{align}
    &\hat{H}_0^\text{MS-CASPT2} = \sum_{M} |M\rangle\langle M| \hat{f} |M\rangle\langle M| + \hat{Q} \hat{f} \hat{Q} + E_s
\\  &\hat{H}_0^\text{XMS-CASPT2} = \hat{P} \hat{f} \hat{P} + \hat{Q} \hat{f} \hat{Q} + E_s
\\  &\hat{f} = \sum_{xy}f_{xy} \hat{E}_{xy}
\\  &f_{xy} = h_{xy} + \sum_{zw} d_{zw}^{(0)\text{sa}} \left[ (xy|zw) - (xw|zy) \right]
\end{align}
using with a projection to the reference space
\begin{align}
    \hat{P} &= \sum_{M} |M\rangle\langle M| 
\\  \hat{Q} &= 1 - \hat{P}
\end{align}
and state-averaged first-order density matrix
\begin{align}
    d_{zw}^{(0)\text{sa}} &= \frac{1}{N_\mathrm{st}}\sum_N \langle N | \hat{E}_{zw} | N \rangle
,\end{align}
where $N_\mathrm{st}$ is the number of states included.  
$E_s$ is a vertical shift parameter.  

The working equations for single-state relativistic fully internally contracted CASPT2 and multi-reference configuration interaction (MRCI) 
were reported by one of us.\cite{shiozaki15jctc}
The working equations for the residual, source, and norm terms were implemented using the code-generator SMITH3\cite{smith3}
and interfaced to the XMS-CASPT2 architecture using the procedure outlined in Ref.~\onlinecite{vlaisavljevich16jctc}.   

The effective Hamiltonian is constructed including shift corrections for all elements:
\begin{align}
\begin{split}
    H_{LL^\prime}^\text{eff} &= H_{LL^\prime}^\text{ref} 
    + \frac{1}{2} \sum_M \left( \langle\tilde{M}|\hat{T}^\dagger_{LM}\hat{H}|\tilde{L}^\prime\rangle + \langle\tilde{L}|\hat{H}\hat{T}_{L^\prime M}|\tilde{M}\rangle \right)
\\  &- E_s \sum_{MN} \langle \tilde{M}|\hat{T}^\dagger_{LM}\hat{T}_{LN}|\tilde{N}\rangle
,\end{split}
\end{align}
the eigenvalues of which correspond to our XMS-CASPT2 second-order energies.  
\begin{align}
    \sum_M H_{LM}^\text{eff} R_{MP} &= R_{LP} E_P^{MS}
\end{align}

Dyall proposed an alternative formulation of multi-reference perturbation theory, where 2-electron interactions among active orbitals are 
fully included in the zeroth-order Hamiltonian, which he called the CAS/$A$ operator.\cite{dyall95jcp}  
\begin{align}
    \hat{H}_0^\text{CAS/$A$} &= \hat{P} \hat{f}^{(A)} \hat{P} + \hat{Q} \hat{f}^{(A)} \hat{Q} + E_s
\\\begin{split}
  \hat{f}^{(A)} &= \sum_{xy}f_{xy} \hat{E}_{xy} - \sum_{rs}f_{rs}\hat{E}_{rs}
\\&+\sum_{rs}h_{rs}\hat{E}_{rs} + \frac{1}{2}\sum_{rs,xy} v_{rs,xy}\hat{E}_{rs,xy}
  \end{split}
\end{align}
Using only the diagonal blocks of Dyall's zeroth-order Hamiltonian allows us to run 
partially contracted NEVPT2.\cite{angeli01jcp,angeli02jcp}
\begin{align}
    \hat{H}_0^\text{NEVPT2} &= \hat{P} \hat{f}^{(A_D)} \hat{P} + \hat{Q} \hat{f}^{(A_D)} \hat{Q} + E_s
\\\begin{split}
    \hat{f}^{(A_D)} &= \sum_{ij}f_{ij} \hat{E}_{ij} + \sum_{ab}f_{ab}\hat{E}_{ab}
\\&+\sum_{rs}h_{rs}\hat{E}_{rs} + \frac{1}{2}\sum_{rs,xy} v_{rs,xy}\hat{E}_{rs,xy}
  \end{split}
\end{align}

\subsection{Mapping to Pseudospin Hamiltonian}

For complexes with spin larger than 1, the energies alone are insufficient to define the ZFS parameters, 
so the \emph{ab initio} states must be mapped to a model Hamiltonian.\cite{maurice09jctc,chibotaru12jcp}  
In order to determine the spin Hamiltonian parameters, we require a mapping from the eigenfunctions of the Hamiltonian 
to a suitable basis of effective or fictitious spin states.  
We first compute the matrix elements of the magnetic moment operator
\begin{align}
    \hat{\boldsymbol{\mu}} &= -\mu_B \sum_j \left[  \frac{1}{2}  g_e \hat{\boldsymbol{\Sigma}}_j
     -i \mathbf{r}_j \times \boldsymbol{\nabla}_j \right]
\end{align}
using the four-component analogue of the Pauli spin operator,
\begin{align}
   \hat{\boldsymbol{\Sigma}} &= 
    \begin{pmatrix}
    \boldsymbol{\sigma} & 0 \\ 0 & \boldsymbol{\sigma}
    \end{pmatrix} 
,\end{align}
where $\boldsymbol{\sigma}$ contains the three standard 2-component Pauli spin matrices.\cite{bauke14pra}   
We build the matrix representation of $\hat{\mu}_{z^\prime}$ 
within the selected manifold of states and diagonalize it.  By convention, 
$z^\prime$ is chosen to be one of the main magnetic axes of the complex, which can be determined as described by 
Chibotaru and Ungur.\cite{chibotaru12jcp,chibotaru13acp} 

The column eigenvectors of $\boldsymbol{\mu}_{z^\prime}$ give us the appropriate combinations of Hamiltonian eigenfunctions 
that should be put in correspondence of pseudospin eigenfunctions, with one complication.  The phases of the column 
eigenvectors are arbitrary, while certain phase restrictions on the pseudospin eigenfunctions must be enforced.  
The first requirement is time-reversal symmetry, the restoration of which is considered in the Supporting Information.  
Apart from time-reversal symmetry, the relative phase between pseudospin eigenfunctions is in principle arbitrary; however, 
we must ensure that the relative phase of true angular momentum matrix elements matches that of the pseudospin operators.  
The conventional choice is that which makes the Clebsch--Gordon coefficients real;\cite{cohentannoudji77v2}
it can be enforced in our context by choosing a phase transformation such that all the 
$\langle m_s | \hat{\mu}_{x^\prime} | m_s+1 \rangle$ matrix elements are real and positive.  
This requirement ensures that the pseudospin reduces to the true spin $\hat{\mathbf{S}}$ at the limit of 
zero spin--orbit and spin--spin coupling and leads to consistent behavior under rotation of the coordinate system.  

\subsection{Extracting the Zero-Field Splitting Parameters}

The numerical pseudospin Hamiltonian is constructed from eigenstates of the \emph{ab initio} Hamiltonian.  
\begin{align}
    &\left(H_\mathrm{diag}\right)_{ij} = \delta_{ij} E_i  \label{Hdiag}
\\  &\mathbf{H}_\mathrm{ZFS} = \mathbf{U}^{\prime -1} \mathbf{H}_\mathrm{diag} \mathbf{U}^\prime \label{H1}
\end{align}
where $E_i$ is the energy of state $i$, and $\mathbf{U}^\prime$ contains the phase-adjusted eigenvectors of $\mu_{z^\prime}$ as discussed above 
and further in the Supporting Information.
When the pseudospin Hamiltonian is built up in terms of extended Stevens operators, we have in matrix form
\begin{align}
    \mathbf{H}_\mathrm{ZFS} &= \sum_{kq} B_k^q \mathbf{O}_k^q \label{H2}
.\end{align}
The extended Stevens operators are built from pseudospin operators; for ZFS, the nonzero contributions 
will have even $k$ between $2$ and $2\tilde{S}$, while $q$ runs from $-k$ to $k$.  The general algorithm for generation of the 
extended Stevens operators has been published by Ryabov\cite{ryabov99jmr,ryabov09amr} and is concisely 
recapitulated in the Supporting Information.   

Setting Eqs.~\eqref{H1} and \eqref{H2} equal leads to a set of linear equations for the coefficients $B_k^q$.  
The second-order contributions are more often expressed in terms of the $D$-tensor
\begin{align}
    \mathbf{H}_\mathrm{ZFS}^{(2)} &= \sum_q B_2^q \mathbf{O}_2^q = \tilde{\mathbf{S}} \cdot \mathbf{D} \cdot \tilde{\mathbf{S}}
,\end{align}
with the parameters
\begin{align}
\begin{split}
    &D_{xx} = -B_2^0 + B_2^2
\\  &D_{zz} = 2B_2^0
\\  &D_{zx} = D_{xz} = \frac{1}{2} B_2^1
\end{split}
\begin{split}
    &D_{yy} = -B_2^0 - B_2^2
\\  &D_{xy} = D_{yx} = B_2^{-2}
\\  &D_{yz} = D_{zy} = \frac{1}{2} B_2^{-1}
\end{split}
\end{align}
and the conventional $D$ and $E$ parameters derived from the diagonalized form of $\mathbf{D}$.  

For dynamically correlated calculations, we approximate the mapping to pseudospin states by using rotated reference CASSCF 
reference functions to determine the magnetic moment matrices, 
\begin{align}
    \boldsymbol{\mu}_{PQ} &= \sum_{MN} \langle \tilde{M} | \hat{\boldsymbol{\mu}} | \tilde{N} \rangle R_{MP}^\ast R_{NQ}
.\end{align}
Using these matrices, the transformation matrix $\mathbf{U}^\prime$ is determined as normal, but the dynamically correlated energies 
are used when building the Hamiltonian in Eq.~\eqref{Hdiag}.  
In this way, we capture the effects of correlation on energy splittings as well as any mixing that occurs in (extended) multi-state methods, 
while avoiding the computation of relaxed density matrices associated with the dynamically correlated methods.  
We consider this approach similar in spirit to the common practice with state interaction methods of using correlated energies along only the 
diagonal elements of the spin--orbit Hamiltonian.\cite{malmqvist02cpl}

\section{Computational Details}

Four-component relativistic calculations were performed with the {\sc BAGEL} electronic structure package\cite{bagel,shiozaki18wires} developed by our group.   
We use spinor basis functions and make use of density fitting with a customized fitting basis set described below.  
Except where otherwise noted, the Dirac--Coulomb relativistic Hamiltonian was used.  
Reference data using the SI-SO approach were computed in MOLCAS\cite{molcas8} and made use of density 
fitting with an atomic compact fitting basis set\cite{aquilante07jcp} generated on the fly.  

There is some difficulty in the use of contracted orbital basis sets with four-component relativistic calculations.  
Part of the basis set incompleteness error stems from the fact that convetional basis sets are optimized using a different Hamiltonian operator, 
but there are additional complications due to restricted kinetic balance.  
Because the RKB condition does not exactly reproduce the relationship between large and small components in the presence of relativistic effects, 
a strongly contracted basis function optimized for the large component will generate a small component basis function that is not optimal.\cite{visscher91ijqc} 
Dyall has made some effort to develop basis sets for fully relativistic calculations 
with different contractions for each of the four components.\cite{dyall98tca,dyall04tca,dyall06tca,dyall10tca}
Currently, our program can only accommodate one set of contraction coefficients in accord with the RKB condition, and in our previous work we 
found Dyall's spin-free basis sets to be inadequate for our needs.\cite{reynolds15pccp}  
The safest approach is to fully decontract the basis set, but many calculations would be intractable with the large primitive basis sets that result.  

In this work, the Atomic Natural Orbitals - Relativistically Core Correlated (ANO-RCC) basis sets\cite{roos05jpca} are used for all atoms.  
In Section~\ref{chalcogenresultsec}, results for a series of small molecules are given for both the uncontracted form 
and the triple-$\zeta$ plus polarization (TZP) contraction of this basis set.  
For four-component calculations on the larger molecules studied in this work, we compromise by using an uncontracted basis on the central metal atom
and contracted basis sets on the ligands, with the set of contractions specified in the appropriate section.  
For the SI-SO comparison data, the TZP contraction is used for central metal atoms, because the atomic mean field integrals used in the SI-SO cannot be accurately computed with a primitive basis set in MOLCAS.\cite{ungur16pc} 

We require a customized auxiliary basis set for density fitting because standard fitting basis sets do not include heavy atoms 
or contain the flexibility needed to capture the Gaunt for full Breit interaction.\cite{kelley13jcp}  To meet these demands, we
have augmented standard basis sets in an even-tempered manner by adding primitive functions with Gaussian exponents increasing by 
a factor of 2.5.  
For H atoms we have simply decontracted the standard TZVPP-JKFIT auxiliary basis set.  
For B, C, N, O, S, Co, and Se, we have decontracted the TZVPP-JKFIT auxiliary basis set, 
and 
we added one tight function to each of the $p$, $d$, and $f$ shells.  
For Te and U, we begin with the decontracted ANO-RCC basis set.  
We add polarization functions by applying the exponents of the highest angular momentum functions to higher shells up to $i$-type
(\emph{i.e.}, for Te the exponents of 3 $g$ functions are copied to the $h$ and $i$ shells, and for U the 3 $h$ exponents are copied to $i$).
We then add several tight functions to each shell, with each one's exponent greater than the last by a factor of 2.5.  
The numbers of functions added are as follows:
3 $s$, 3 $p$, 6 $d$, 7 $f$, 5 $g$, and 1 $h$ for Te, and 4 $s$, 4 $p$, 7 $d$, 8 $f$, 6 $g$, 6 $h$, and 5 $i$ for U.

For dynamically correlated calculations, core orbitals were frozen to reduce the cost of the calculation.  
In some cases, the higher-energy virtual orbitals resulting from basis set decontraction were deleted as well; the energy cutoffs are specified in the relevant results section.  
A vertical shift of 0.2 was used in four-component XMS-CASPT2 and NEVPT2 calculations, with correction terms included in all elements of the 
effective Hamiltonian.\cite{roos95cpl,vlaisavljevich16jctc}  
For CASPT2 + SI-SO, an imaginary shift of 0.2 was used along with the standard 0.25 IPEA shift.\cite{forsberg97cpl,ghigo04cpl}  

All computations were performed at the experimental structures.\cite{huber79,rinehart10jacs,zadrozny13poly,suturina17ic} 
The bond lengths (for diatomics) and molecular coordinates (for larger molecules) are provided 
in the Supporting Information.   
For the larger molecules, heavy atom positions were obtained from the crystal structure, and hydrogen positions were optimized using DFT.  
The DFT optimizations were performed using the Turbomole package.\cite{ahlrichs89cpl,furche14wires}
We used the PBE functional\cite{perdew96prl} along with the def2-TZVPP basis set for all atoms except U, for which we used def-TZVPP with effective core potential.\cite{weigend98cpl,weigend05pccp,dolg89jcp,kuchle94jcp,cao03jcp} 
We also used the resolution of the identity approximation along with the corresponding RI-J auxiliary basis sets.\cite{eichkorn97tca,weigend06pccp}

\section{Results and Discussion}

\subsection{Benchmarking Dynamical Correlation Methods with Chalcogen Diatomics} \label{chalcogenresultsec}

\begin{table*}[tb]
\caption{
   Axial ZFS ($D$) of chalcogen diatomics computed with various methods for an active space of 8 electrons in 6 orbitals.  
   Rhombic splitting ($E$) is zero by symmetry.
   \label{chalcogendata2}
}
\begin{ruledtabular}
\begin{tabular}{ldddddddddddd}
 \multicolumn{1}{c}{Molecule}   && \multicolumn{2}{c}{O$_2$} && \multicolumn{2}{c}{S$_2$} && \multicolumn{2}{c}{Se$_2$} && \multicolumn{2}{c}{Te$_2$} \\
 \multicolumn{1}{c}{States}     && \multicolumn{1}{c}{Triplet} & \multicolumn{1}{c}{6 lowest} && \multicolumn{1}{c}{Triplet} & \multicolumn{1}{c}{6 lowest} && \multicolumn{1}{c}{Triplet} & \multicolumn{1}{c}{6 lowest} && \multicolumn{1}{c}{Triplet} & \multicolumn{1}{c}{6 lowest} \\ \hline
    \\ \multicolumn{13}{c}{TZP contraction of ANO-RCC} \\
  CASSCF      &&   3.8 &   3.0 &&  24.4 &  19.9 &&  505.6 &  434.5 &&  2067.4 &  1896.0 \\
  XMS-CASPT2  &&   1.0 &   2.7 &&   8.2 &  19.7 &&  234.4 &  412.6 &&  1428.6 &  1913.8 \\
  NEVPT2      &&   2.7 &   2.9 &&  20.1 &  20.0 &&  424.3 &  422.5 &&  1931.3 &  1952.1 \\
  MRCI+Q      &&   2.7 &   2.9 &&  22.5 &  20.7 &&  470.9 &  440.6 &&  2052.4\footnotemark[1] &  1955.6\footnotemark[1] \\

  \\ \multicolumn{13}{c}{Decontracted ANO-RCC} \\
  CASSCF      &&   3.8 &   3.0 &&  27.9 &  22.7 &&  558.7 &  480.3                &&  2088.6                 &  1905.9 \\
  XMS-CASPT2  &&   1.1 &   2.9 &&   9.9 &  23.9 &&  278.6\footnotemark[2] &  489.7\footnotemark[2]&&  1401.1\footnotemark[1]^,\footnotemark[2] &  1944.3\footnotemark[1]^,\footnotemark[2] \\
  NEVPT2      &&   2.9 &   3.0 &&  24.0 &  24.5 &&  497.0\footnotemark[2] &  502.7\footnotemark[2]&&  1957.1\footnotemark[1]^,\footnotemark[2] &  1994.8\footnotemark[1]^,\footnotemark[2] \\
\\
  Experiment  &&  \multicolumn{2}{c}{ 4.0\footnotemark[3]}   &&  \multicolumn{2}{c}{23.5\footnotemark[4]}   &&  \multicolumn{2}{c}{510.0\footnotemark[5]}   &&  \multicolumn{2}{c}{1975.0\footnotemark[5]}
\end{tabular}
\end{ruledtabular}
\footnotetext[1]{Additional semi-core orbitals frozen.}
\footnotetext[2]{Virtual orbitals with energy above 100 Hartrees were deleted.}
\footnotetext[3]{Ref.~\citenum{tinkham55pr}}
\footnotetext[4]{Refs.~\citenum{pickett79jms} and \citenum{wayne74mp}}
\footnotetext[5]{Ref.~\citenum{huber79}}
\end{table*}

Table~\ref{chalcogendata2} presents ZFS results for a series of chalcogen diatomics using the methods reported in the previous sections.
The diatomics in this series have been previously used to benchmark other methods by Rota \emph{et al.}\cite{rota11jcp}
For dynamically correlated calculations, deep core orbitals were frozen (2 for S$_2$, 10 for Se$_2$, and 18 for Te$_2$).  
As indicated in the table, an additional 18 semi-core orbitals were frozen in some Te$_2$ calculations to reduce the cost.  
These benchmark tests were performed with the MS-MR parametrization.  

We have also tested the influence of basis set contraction upon ZFS results, because ANO-RCC basis contractions are
not optimized for the four-component treatment of relativistic effects.  
Due to cost constraints, MRCI+Q results are included only for the contracted basis set, 
and in some cases high-energy virtual orbitals were deleted.  
In most cases, the results with the contracted basis set are further from the experimental values, with errors of 10--20\% in the cases of S$_2$ and Se$_2$, 
and smaller errors for O$_2$ and Te$_2$.  
This suggests that in the pursuit of high-quality results without the large costs associated with decontraction, there is use for further development of 
basis sets suitable for four-component relativistic calculations.  For the time being, we 
prefer using uncontracted basis sets and compromise when necessary by contracting basis sets on ligands but not spin centers.  

The XMS-CASPT2 results in Table~\ref{chalcogendata2} show an extremely strong dependence upon the choice of states to include in the calculation, 
with large errors introduced when excluding some low-lying excited singlets.  
The choice of which excited states to include affects the calculation primarily in two ways: through the 
orbital optimization in state-averaged CASSCF, and through the zeroth-order Hamiltonian which uses a state-averaged density matrix.  
Table~S2 in the Supporting Information separates these effects by correlating the first three or six states using both sets of orbitals, and clearly 
shows that the zeroth-order Hamiltonian is the main source of error; the change to the state-averaged orbitals alters the XMS-CASPT2 splitting by a few percent only.  
It is unsurprising, then, that the strong dependence on state averaging is not observed in the other methods; Dyall's Hamiltonian used by 
NEVPT2 replaces the active-active part of the state-averaged Fock matrix with the full Hamiltonian, and MRCI+Q does not use it at all.  

MRCI+Q is unsurprisingly the most reliable method tested here, although for large problems it will be intractable.  
When a more approximate dynamically correlated method is needed, NEVPT2 is preferred due to its relatively low sensitivity to the set of states included, 
although for this series the accuracy of CASSCF is competitive with the perturbative methods, in comparison to experimental values.  

We briefly mention one other complication which has been observed.   
For calculations with a very minimal active space (\emph{e.g.}, 2 electrons in 2 orbitals with all six states included), it is possible for states 
which are nondegenerate under $\hat{H}$ to become degenerate under $\hat{H}_0$ and, therefore, to freely mix in the XMS rotation.  
In such cases, if a vertical shift is used with only the diagonal correction terms, unphysical breaking of Kramers degeneracy will result, 
but if no shift is used or we correct the entire effective Hamiltonian, degeneracy is restored in the final multi-state mixing step.

\subsection{Zero-Field Splitting of Pseudotetrahedral Cobalt(II) $(\boldsymbol{X}$Ph$)_{\boldsymbol{4}}$ Complexes}

\begin{figure*}[t]
\includegraphics[width=7.0in]{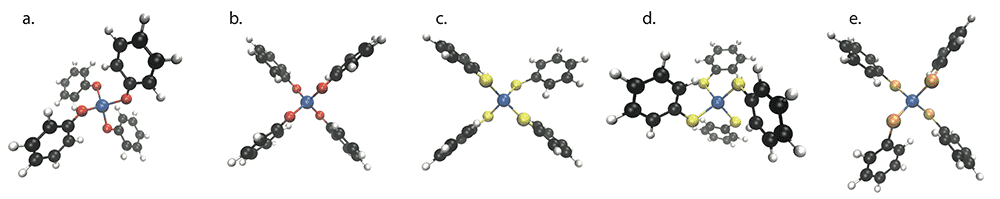}
\caption{Geometry of the Co$(X$Ph$)_4$ complexes tested.  
(a.) Co(OPh)$_4^{2-}$ crystallized with 2 PPh$_4^+$ and 2 CH$_3$CN.\cite{zadrozny13poly}  
(b.) Co(OPh)$_4^{2-}$ crystallized with PPh$_4^+$ and K$^+$.\cite{zadrozny13poly}  
(c.) Co(SPh)$_4^{2-}$ crystallized with 2 PPh$_4^+$.\cite{suturina17ic}
(d.) Co(SPh)$_4^{2-}$ crystallized with 2 N(CH$_2$CH$_3)_4^+$.\cite{suturina17ic}
(e.) Co(SePh)$_4^{2-}$ crystallized with 2 PPh$_4^+$.\cite{zadrozny13poly}
}
\label{cox4pics}
\end{figure*}

To gauge the performance of relativistic correlation methods for mononuclear transition metal complexes, 
we look to the series of pseudotetrahedral cobalt(II) $(X$Ph$)_4$ complexes with $X$ = O, S, Se
whose structures are shown in Figure~\ref{cox4pics}.  Molecule (c), (PPh$_4$)$_2$Co(SPh)$_4$ was one of the first 
field-free transition metal single-ion magnets to be discovered;\cite{zadrozny11jacs} the analogous complexes here 
have been characterized both experimentally and theoretically to probe the magneto-structural correlations 
present.\cite{zadrozny13poly,suturina15ic,suturina17ic} 
Structures (a) and (b) as well as (c) and (d) differ only in the counter ion and solvent present in the crystal, 
which dramatically alters both the conformation and ZFS of the anion.  

CASSCF-level ZFS parameters computed for the whole series are compared along with experimental values 
in Table~\ref{cox4data}.  
A mixed basis set is used, with the central cobalt atom decontracted in four-component results, the 
TZP contraction for chalcogen atoms, the double-$\zeta$ plus polarization (DZP) contraction for carbon atoms, and 
double-$\zeta$ (DZ) for hydrogen atoms.  
Agreement with experiment is reasonably good for this series, with qualitative agreement throughout.  
The major exception is the Dirac-CASSCF results with only the ground quartet included in the state-averaging 
procedure; in most cases more accurate results are obtained when the orbitals 
are balanced by including all $d^5$ excitations (10 quartets and 40 doublets).  
The choice of whether to include a second, correlating set of $d$-orbitals is of lesser importance, typically altering 
the observed splitting by a few wavenumbers or less.  

\begin{table*}[tb]
\centering
\caption{ZFS of the pseudotetrahedral cobalt(II) $(X$Ph$)_4$ series at the CASSCF level of theory.   Energies are given in cm$^{-1}$.  
   Calculations are based on an active space of 7 electrons in either 5 or 10 active $d$-orbitals.  
   \label{cox4data}
}
\begin{ruledtabular}
\begin{tabular}{clrdrdrdl}
  \multirow{2}{*}{Molecule} & \multirow{2}{*}{Active Space} & \multicolumn{2}{c}{Dirac Ham. } &  \multicolumn{2}{c}{Dirac Ham.} & \multicolumn{2}{c}{Scalar + SI-SO}
                                              & \multirow{2}{*}{Experiment} \\
             & & \multicolumn{2}{c}{ground quartet} &  \multicolumn{2}{c}{full $d^5$} & \multicolumn{2}{c}{(full $d^5$)}
                                                                                      \\ \hline
                     & \multirow{2}{*}{7 e$^-$, 5  orb.} & $D = $& -9.5 & $D = $&-13.6 & $D = $&-13.5 & $         $ \\
  Co(OPh)$_4^{2-}$   &                                   & $E = $&  1.4 & $E = $&  2.1 & $E = $&  2.1 & $D = -11.1(3)\footnotemark[1]$ \\
  Geometry (a)       & \multirow{2}{*}{7 e$^-$, 10 orb.} & $D = $&-10.1 & $D = $&-13.5 & $D = $&-12.6 & $(E = 0.0)$ \\
                     &                                   & $E = $&  1.4 & $E = $&  1.9 & $E = $&  1.8 & $         $ \\ \hline
                     & \multirow{2}{*}{7 e$^-$, 5  orb.} & $D = $&-15.1 & $D = $&-24.6 & $D = $&-24.5 & $         $ \\
  Co(OPh)$_4^{2-}$   &                                   & $E = $&  0.0 & $E = $&  0.0 & $E = $&  0.0 & $D = -23.8(2)\footnotemark[1]$ \\
  Geometry (b)       & \multirow{2}{*}{7 e$^-$, 10 orb.} & $D = $&-16.2 & $D = $&-22.5 & $D = $&-22.5 & $(E = 0.0)$ \\
                     &                                   & $E = $&  0.0 & $E = $&  0.0 & $E = $&  0.0 & $         $ \\ \hline
                     & \multirow{2}{*}{7 e$^-$, 5  orb.} & $D = $&-31.1 & $D = $&-58.3 & $D = $&-58.3 & $         $ \\
  Co(SPh)$_4^{2-}$   &                                   & $E = $&  1.6 & $E = $&  1.6 & $E = $&  1.6 & $D = -62(1)\footnotemark[1], -55(1)\footnotemark[2]$ \\
  Geometry (c)       & \multirow{2}{*}{7 e$^-$, 10 orb.} & $D = $&-36.2 & $D = $&-58.0 & $D = $&-58.0 & $E = 0(2)\footnotemark[2]$ \\
                     &                                   & $E = $&  1.7 & $E = $&  1.5 & $E = $&  1.5 & $         $ \\ \hline
                     & \multirow{2}{*}{7 e$^-$, 5  orb.} & $D = $& 10.9 & $D = $& 16.9 & $D = $& 16.9 & $         $ \\
  Co(SPh)$_4^{2-}$   &                                   & $E = $&  1.7 & $E = $&  2.5 & $E = $&  2.5 & $D = 11(2)\footnotemark[2]$ \\
  Geometry (d)       & \multirow{2}{*}{7 e$^-$, 10 orb.} & $D = $& 11.9 & $D = $& 15.3 & $D = $& 15.3 & $E = 2.0(4)\footnotemark[2]$ \\
                     &                                   & $E = $&  1.8 & $E = $&  2.3 & $E = $&  2.3 & $         $ \\ \hline
                     & \multirow{2}{*}{7 e$^-$, 5  orb.} & $D = $&-38.7 & $D = $&-80.4 & $D = $&-80.4 & $         $ \\
  Co(SePh)$_4^{2-}$  &                                   & $E = $&  1.5 & $E = $&  1.2 & $E = $&  1.2 & $D = -83(1)\footnotemark[1]  $ \\
  Geometry (e)       & \multirow{2}{*}{7 e$^-$, 10 orb.} & $D = $&-45.3 & $D = $&-80.3 & $D = $&-80.3 & $(E = 0.0)$ \\
                     &                                   & $E = $&  1.7 & $E = $&  1.0 & $E = $&  1.0 & $         $ \\
\end{tabular}
\end{ruledtabular}
\footnotetext[1]{Ref.~\citenum{zadrozny13poly}}
\footnotetext[2]{Ref.~\citenum{suturina17ic}}
\end{table*}

Four-component NEVPT2 calculations have also been performed on structures (a), (c), and (e) of this series; 
the results are summarized in Table~\ref{cox4datashort}.  
Due to cost constraints, it was necessary to reduce the basis set combination used: for these data, the basis sets 
for cobalt and hydrogen are the same as above, but the DZP contraction was used for chalcogen atoms and the DZ contraction for carbon atoms.  
All results in this table use an active space of 7 electrons in 5 active orbitals.  
For Dirac-NEVPT2, the MS-MR contraction is used while correlating the four lowest states, but using orbitals that were obtained by finding a 
minimax state-averaged energy for all excitations of $d$-electrons.  
The number of frozen core orbitals is 37, 53, and 69 for structures (a), (c), and (e), respectively. 
89 high-energy virtual orbitals were deleted, corresponding to an energy cutoff threshold of 11 $E_\mathrm{h}$
to avoid splitting up near-degenerate orbitals at 10 $E_\mathrm{h}$.

We used the MS-MR internal contraction scheme in this and the previous section.  It is tempting to switch to the 
SS-SR parametrization, which would reduce the memory cost by roughly a factor of $N_\mathrm{st}$ and the timing cost by a greater amount.   
Some tests have been performed using truncated models for this molecular series; the details are presented in the Supporting Information.  
We found that reducing the expansion set to the SS-SR scheme introduced 
errors in the splitting energies of several percent, and that the ratio $|E/D|$ was much more sensitive.   
Therefore, we prefer the MS-MR parametrization when possible, while noting that the errors introduced by SS-SR in the 
energy spacings were moderate.  

We see corrections due to dynamical correlation of 2 to 20 cm$^{-1}$, slightly larger than are observed in the CASPT2 + SI-SO reference data.  
The basis set reduction necessary to make the NEVPT2 calculations tractable introduced errors of up to 3.6 cm$^{-1}$ at the Dirac-CASSCF level, showing 
somewhat higher basis set sensitivity than the SI-SO results; one might expect a somewhat greater basis set incompleteness error in the correlated data, which 
would be consistent with the results in Table~\ref{chalcogendata2} for smaller test systems.  For structure (a), the Dirac-NEVPT2 best reproduces the 
experimental findings, but for the other two complexes it appears to overcorrect.  

\begin{table*}[tb]
\centering
\caption{ZFS for a subset of the pseudotetrahedral cobalt(II) $(X$Ph$)_4$ series.   Energies are given in cm$^{-1}$.  
   Calculations are based on an active space of 7 electrons in 5 active $d$-orbitals and with a smaller basis set than in Table~\ref{cox4data}, 
   as discussed in the main text.  
   \label{cox4datashort}
}
\begin{ruledtabular}
\begin{tabular}{crdrdrdrdl}
  \multirow{2}{*}{Molecule} & \multicolumn{4}{c}{CASSCF} & \multicolumn{2}{c}{NEVPT2} & \multicolumn{2}{c}{CASPT2} &\multirow{3}{*}{Experiment} \\
             & \multicolumn{2}{c}{Dirac Ham. } & \multicolumn{2}{c}{Scalar + SI-SO}
               & \multicolumn{2}{c}{Dirac Ham.} & \multicolumn{2}{c}{Scalar + SI-SO} \\ \hline
 \multirow{2}{*}{Co(OPh)$_4^{2-}$ (a)   }  & $D = $&-13.9 & $D = $&-13.9 & $D = $& -11.3& $D = $&-13.1 &  $D = -11.1(3)\footnotemark[1]$                \\
                                           & $E = $&  2.1 & $E = $&  2.1 & $E = $&   1.5& $E = $&  1.7 & ($E =   0.0$)                  \\ \hline
 \multirow{2}{*}{Co(SPh)$_4^{2-}$ (c)  }   & $D = $&-61.9 & $D = $&-58.8 & $D = $&-43.3 & $D = $&-53.1 &  $D = -62(1)\footnotemark[1], -55(1)\footnotemark[2]$ \\
                                           & $E = $&  1.6 & $E = $&  1.6 & $E = $&  1.5 & $E = $&  1.1 &  $E =   0(2)\footnotemark[2] $                  \\ \hline
 \multirow{2}{*}{Co(SePh)$_4^{2-}$ (e) }   & $D = $&-83.1 & $D = $&-80.9 & $D = $&-63.8 & $D = $&-72.4 &  $D = -83(1)\footnotemark[1]$                  \\
                                           & $E = $&  1.2 & $E = $&  1.0 & $E = $&  1.3 & $E = $&  0.7 & ($E =   0.0)$                  \\
\end{tabular}
\end{ruledtabular}
\footnotetext[1]{Ref.~\citenum{zadrozny13poly}}
\footnotetext[2]{Ref.~\citenum{suturina17ic}}
\end{table*}

\subsection{Zero-Field Splitting of a Uranium(III) Pyrazole Complex}

\begin{figure}[t]
\includegraphics[width=2.25in]{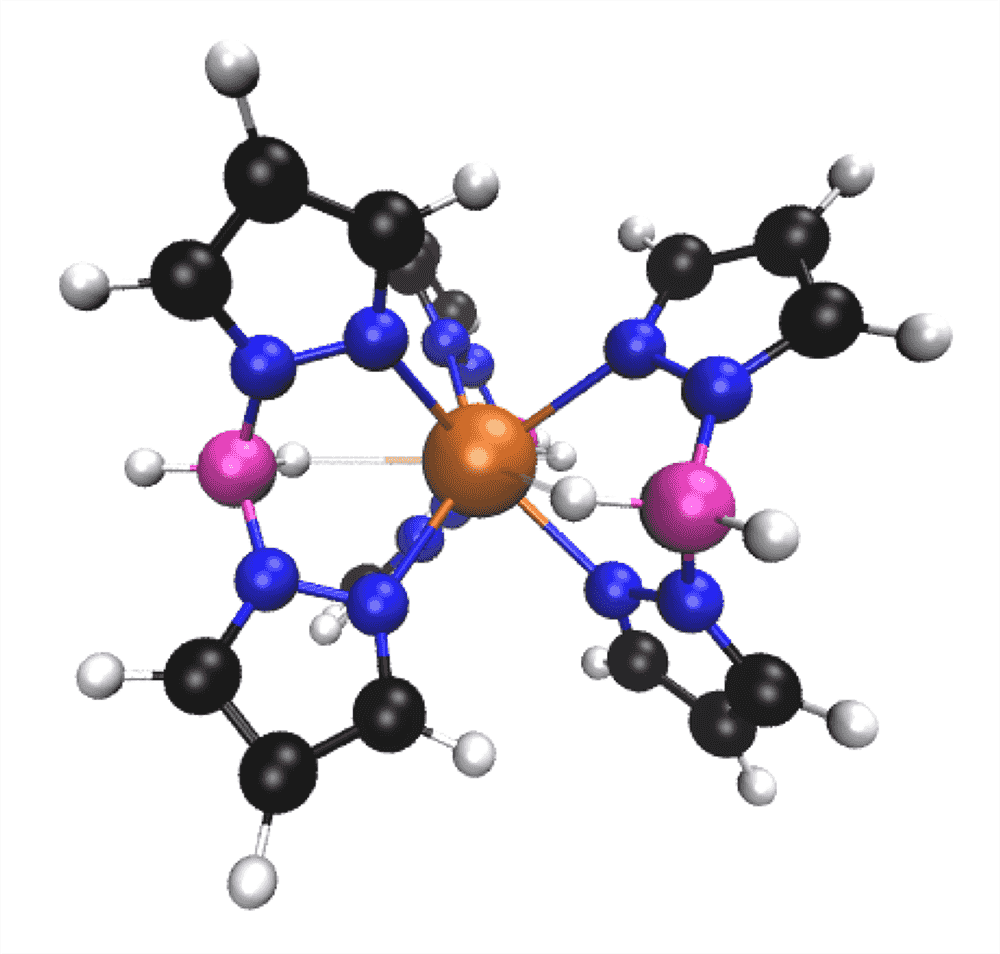}
\caption{Geometry of U(H$_2$BPz$_2$)$_3$.\cite{rinehart10jacs} } 
\label{uraniumpic}
\end{figure}

To test the performance of these methods for magnetic complexes with very heavy elements, we have 
computed the ZFS of the ground $J=\tfrac{9}{2}$ multiplet in the uranium(III) dihydrobispyrazolylborate complex U(H$_2$BPz$_2$)$_3$,
shown by Long and co-workers to act as a single-ion magnet in magnetically dilute samples.\cite{rinehart10jacs,meihaus11ic}
The structure is depicted in Figure~\ref{uraniumpic}, and results are presented in Table~\ref{uraniumdata}.
Previous theoretical calculations have been carried out with the SI-SO approach by Spivak \emph{et al.}\cite{spivak17jpca}
and with a semi-empirical crystal-field approach by Baldovi \emph{et al.}\cite{baldovi13cs}  
The results reported by Spivak \emph{et al.} show a somewhat smaller splitting than that in Table~\ref{uraniumdata},  
which might result from a different choice of molecular geometry; further details are discussed in the Supporting Information.  

The results in Table~\ref{uraniumdata} were computed with either the decontracted basis or TZP contraction for the uranium atom as appropriate, 
the DZP basis set for boron, carbon, and nitrogen atoms, and the DZ contraction for hydrogen atoms.  
An exception is the Dirac-NEVPT2 result, for which the polarization functions were not used for main group elements;
one column also shows CASSCF results with the same basis set.   
The active space in all cases is 3 active electrons in the 7 $f$-orbitals.  
Orbital optimizations were perfomed for the minimax state-averaged energy for all excitations within the $f$-orbitals.  
NEVPT2 calculations included the lowest 10 states and used the SS-SR parametrization; 72 core orbitals were frozen, and 
160 virtual orbitals were deleted, corresponding to an energy cutoff of 10 $E_\mathrm{h}$.  

We computed an excitation energy of 161.2 cm$^{-1}$ with Dirac-CASSCF and 103.3 cm$^{-1}$ with NEVPT2, 
somewhat lower than the value of 
230 cm$^{-1}$ predicted by a semi-empirical crystal-field approach.\cite{baldovi13cs}
We obtain qualitatively similar results with Dirac-CASSCF vs. CASSCF + SI-SO.  
The Breit correction at the CASSCF level was determined to be only a few wavenumbers, 
smaller than other errors relating to electron correlation and the molecular geometry dependence.  

The excitation spectrum of U(H$_2$BPz$_2$)$_3$ is not available for comparison.  
The closest analogue we have is a measurement of the magnetic relaxation barrier, which was found to be 
16 cm$^{-1}$ in a magnetically dilute sample.\cite{meihaus11ic}  
All theoretical methods examined here give a first excited state at least several times higher.  
There are several competing relaxation mechanisms in this molecule,\cite{meihaus11ic} 
so the observed barrier is not directly related to the energy splitting computed in this work.  

The CASPT2 + SI-SO approach requires computing the second-order correlation energy for a large number of excited 
states, and there are practical limits to how many states can be included.  We have truncated our 
CASPT2 space at 21 quartets and 11 doublets, out of 35 quartets and 112 doublets accessible by exciting within the $f$-orbital manifold,
which is the same cutoff used in the earlier study.\cite{spivak17jpca}
As shown in Table~\ref{uraniumdata}, truncating the coupling of CASSCF states at the same level shifts the 
energy splittings by 70-160 cm$^{-1}$, so the aggressive truncation needed for large active spaces 
comes at a substantial accuracy cost for ZFS calculations.  
One of the advantages of the four-component approach is that spin--orbit effects are included from the start and can be 
captured without computing many excited states, so it avoids this particular problem.  
Of course, there is a tradeoff; the large memory costs associated with the four-component NEVPT2 
required us to make compromises to limit the number of orbitals included in the correlation,
although in this case the errors associated with reducing the basis set are manageable, at least at the CASSCF-level.  
For this molecule, the Dirac-NEVPT2 correction to the enery splittings is of comparable magnitude but opposite sign
relative to the CASPT2 + SI-SO reference.  

\begin{table*}[tb]
\centering
\caption{Low-lying energy spectrum of U(H$_2$BPz$_2$)$_3$.  Energies are given in cm$^{-1}$.  
   Calculations are based on an active space of 3 electrons in the 7 active $f$-orbitals.  
   \label{uraniumdata}
}
\begin{ruledtabular}
\begin{tabular}{lddddddddd}
  \multirow{2}{*}{Eigenstates } & \multicolumn{4}{c}{Dirac Hamiltonian} & \multicolumn{3}{c}{State-interaction with Spin-Orbit} & \multicolumn{1}{c}{Crystal field}\\
             & \multicolumn{1}{c}{CASSCF} &  \multicolumn{1}{c}{Breit corr.} & \multicolumn{1}{c}{CASSCF\footnotemark[1]} & \multicolumn{1}{c}{NEVPT2\footnotemark[1]} &  \multicolumn{1}{c}{CASSCF (full)} &  \multicolumn{1}{c}{CASSCF (trunc.)} &  \multicolumn{1}{c}{CASPT2 (trunc.)} & \multicolumn{1}{c}{Semi-empirical\footnotemark[2]} \\ \hline
States  1 - 2  &   0.0  &   0.0 &   0.0 &    0.0  &   0.0 &   0.0  &   0.0  &   0         \\
States  3 - 4  & 161.2  &   1.3 & 163.3 &  103.3  & 165.4 &  96.4  & 154.2  & 231         \\
States  5 - 6  & 407.5  &  -2.3 & 412.3 &  425.4  & 406.9 & 286.8  & 275.0  & 491         \\
States  7 - 8  & 524.1  &   0.2 & 525.0 &  503.9  & 511.4 & 392.3  & 397.9  & 609         \\
States  9 - 10 & 644.4  &   2.8 & 651.0 &  570.1  & 611.2 & 452.0  & 482.4  & 842         \\
\end{tabular}
\end{ruledtabular}
\footnotetext[1]{Basis set reduced by removing polarization functions from the B, C, and N atoms.}
\footnotetext[2]{Calculated from crystal field parameters reported by Baldovi \emph{et al.}\cite{baldovi13cs}}
\end{table*}

\section{Conclusions}

We have developed a computational tool for the determination of ZFS parameters in relatively 
large molecules containing strong relativistic effects.  
Specifically, we report the development of (extended) multistate implementations of four-component relativistic CASPT2, 
NEVPT2, and MRCI as well as a mapping from the four-component wavefunctions to a pseudospin Hamiltonian for the 
extraction of ZFS parameters.  
We observed a surprisingly large dependence upon the state-averaging procedure used to obtain the orbitals, which is 
troublesome for XMS-CASPT2 but reduced in CASSCF, NEVPT2, and MRCI+Q.  
Benchmark calculations performed on the series of chalcogen diatomics and pseudotetrahedral Co$(X$Ph$)_4$ complexes 
showed similar results to the less costly scalar relativistic + SI-SO approach, suggesting that for light elements 
the fully relativistic methods can serve as a useful benchmark but are not always needed.  
Tests on the actinide single-molecule magnet U(H$_2$BPz$_2$)$_3$ showed larger effects, indicating that 
for very heavy metals the simultaneous treatment of relativity and electron correlation, as well as avoidance of 
state truncation sometimes required for SI-SO, can be helpful when attempting high-accuracy calculations. 
The methods reported in this manuscript are publicly available in the {\sc BAGEL} program package\cite{bagel} and distributed under the GNU General Public License.  

\section{Associated Content}
\subsection{Supporting Information}

The following content is available in the Supporting Information:  
Discussion of the theory and working equations used in the implementation of the mapping to the pseudospin Hamiltonian; 
ZFS results for chalcogen diatomics with a reduced active space, including an investigation to separate the 
effects of state-averaged orbitals and state-averaged Fock operator on dynamically correlated results;
comparison of ZFS results using the SS-SR and MS-MR parametrizations for truncated models of the Co($X$Ph)$_4^{2-}$ series;
discussion of the molecular geometry of U(H$_2$BPz$_2$)$_3$ and the effects of optimization of hydrogen positions;
xyz coordinates of the molecules used for testing.

\section{Acknowledgments}

The authors thank Scott Coste and Danna Freedman for helpful discussions.  
RDR has been supported by the DOD National Defense Science and Engineering Graduate (NDSEG) Fellowship, 32 CFR 168a.
TS has been supported by the NSF CAREER Award (CHE-1351598). 
TS is an Alfred P. Sloan Research Fellow.
Some of the computations were performed with the aid of the ERDC DSRC high-performance computing resources (AFOSR40403702).

\end{document}